\documentstyle[12pt]{article}
\makeatletter
\@addtoreset{equation}{section}
\makeatother

\renewcommand{\baselinestretch}{1.2}
\newcommand{\foot}{\footnote}
\newcommand{\equ}[1]{(\ref{#1})}
\newcommand{\be}{\begin{equation}}
\newcommand{\ee}{\end{equation}}
\newcommand{\eel}[1]{\label{#1}\end{equation}}
\newcommand{\bea}{\begin{eqnarray}}
\newcommand{\eea}{\end{eqnarray}}
\newcommand{\eeal}[1]{\label{#1}\end{eqnarray}}
\newcommand{\beac}{\begin{equation}\begin{array}{rcl}}
\newcommand{\eeacn}[1]{\end{array}\label{#1}\end{equation}}
\newcommand{\qq}{&\qquad &}
\newcommand{\del}{\partial}
\newcommand{\delb}{\bar{\partial}}
\newcommand{\eps}{\epsilon}
\newcommand{\Th}{\hat{T}}
\newcommand{\Gh}{\hat{G}}
\newcommand{\Jh}{\hat{J}}
\newcommand{\fl}{\phi_l}
\newcommand{\pl}{\psi_l}
\newcommand{\bfl}{\bar{\phi}_l}
\newcommand{\bpl}{\bar{\psi}_l}
\newcommand{\gp}{\gamma_{+}}
\newcommand{\gm}{\gamma_{-}}
\newcommand{\bp}{\beta_{+}}
\newcommand{\bm}{\beta_{-}}

\newcommand{\bpm}{\beta_{\pm}}
\newcommand{\non}{\nonumber}
\newcommand{\jg}{\jmath_{ghost}}
\newcommand{\jgp}{\jmath_{ghost+}}
\newcommand{\jgm}{\jmath_{ghost-}}
\newcommand{\Jg}{{\cal J}_{ghost}}
\newcommand{\I}{{\cal I}}
\newcommand{\Ibrs}{{\cal I}_{BRST}}
\newcommand{\la}{{\lambda}}
\newcommand{\lab}{{\bar{\lambda}}}
\newcommand{\psib}{{\bar{\psi}}}
\newcommand{\chib}{{\bar{\chi}}}
\newcommand{\mub}{{\bar{\mu}}}
\newcommand{\nub}{{\bar{\nu}}}
\newcommand{\rhob}{{\bar{\rho}}}
\newcommand{\taub}{{\bar{\tau}}}
\newcommand{\phib}{{\bar{\phi}}}
\newcommand{\Ab}{{\bar{A}}}
\newcommand{\Ah}{{\hat{A}}}
\newcommand{\At}{{\tilde{A}}}
\newcommand{\ka}{{\kappa}}
\newcommand{\etah}{{\hat{\eta}}}
\newcommand{\etat}{{\tilde{\eta}}}

\begin{document}
%
\begin{titlepage}
\topskip0.5cm
\hfill\hbox{CERN-TH.7379/94}\\[-1.4cm]
\flushright{\hfill\hbox{hep-th/9408033}}\\[3.3cm]
\begin{center}{\Large\bf
Superstrings from Hamiltonian Reduction\\[2cm]}{
\Large A. Boresch, K. Landsteiner, W. Lerche, A. Sevrin
\\[1.5cm]}
CERN, Geneva, Switzerland\\
[2.5cm]
\end{center}
\begin{abstract}
In any string theory there is a hidden, twisted superconformal
symmetry algebra, part of which is made up by the BRST current and
the anti-ghost. We investigate how this algebra can be systematically
constructed for strings with $N\!-\!2$ supersymmetries, via quantum
Hamiltonian reduction of the Lie superalgebras $osp(N|2)$. The
motivation is to understand how one could systematically construct
generalized string theories from superalgebras. We also briefly
discuss the BRST algebra of the topological string, which is a doubly
twisted $N\!=\!4$ superconformal algebra.
\end{abstract}
\vfill
\hbox{CERN-TH.7379/94}\hfill\\
\hbox{July 1994}\hfill\\
\end{titlepage}
%
%
\renewcommand{\baselinestretch}{1.3}\large\normalsize
\section{Introduction}

It is known that the BRST structure of (critical and non-critical)
bosonic strings can be characterized in terms of a twisted $N\!=\!2$
superconformal algebra \cite{gase}, that of superstrings in terms of
an $N\!=\!3$ algebra and that of $W_n$-strings in terms of $N\!=\!2$
$W_n$-algebras \cite{blnw}, etc. In terms of a given superconformal
algebra, the nilpotent BRST operator is nothing but one of the
supercharges,
\be
{{\cal Q}_{BRST}}\ \equiv\ G^+_0\ =\ \oint dz\,(cT+\dots)\ ,
\eel{QG}
such that the fundamental relation $T(z)=\{{{\cal Q}_{BRST}},b(z)\}$
is part of an $N\!=\!2$ sub-algebra (the anti-ghost is just the
conjugate supercurrent, $G^-(z)=b(z)$). The above expression
represents a very specific realization of such superconformal
algebras, and it was shown in \cite{blnw,ki} that for ordinary
bosonic and $W_n$-strings, this kind of realization can indeed be
systematically obtained via quantum Hamiltonian reduction from WZW
models based on $s\ell(n|n\!-\!1)$. (Note that there exists a
different, complementary way of characterizing the BRST structure in
terms of superconformal algebras, initiated in \cite{BV}. These BRST
algebras seem to be much harder to obtain from Hamiltonian reduction,
if at this is all possible, because they involve bosonization of the
$\beta,\gamma$ systems.)

Embedding of string theories into WZW models allows to analyze
their BRST structure by Lie algebraic methods, and more
importantly, opens up the possibility of classifying string
theories in terms of superalgebras. Since the Hamiltonian reduction
of $s\ell(2|1)$ yields the ordinary bosonic string, one may expect
that
a very large class extensions of the bosonic string are determined by
the embeddings of $s\ell(2|1)$ into arbitrary superalgebras. However,
what
really makes a string model is certainly not just any superconformal
algebra obtained in this way: in order to define a string theory,
one also must require a particular realization of the algebra, namely
one where one of the supercharges has the form\foot{Note that
$G^+(z)$ will in general explicitly contain the Liouville field or
some matter fields, so that it is not really covariant. However,
since
these fields appear only in total derivative terms, this is
irrelevant for the classification of BRST operators.}~\equ{QG}.

The existence of such realizations is a priori not obvious, and it is
the purpose of this paper to shed some light on this problem of
constructing general string BRST operators from superalgebras. We
will primarily focus on how superstrings with $N\!-\!2$
supersymmetries can
be obtained via quantum Hamiltonian reduction from the superalgebras
$osp(N|2)$. It turns out that in comparison to $W_n$-strings
(associated with $s\ell(n|n\!-\!1)$), the situation is much more
involved here, in that various complications occur. To our knowledge,
some of these complications in the Hamiltonian reduction and their
resolution have so far not been treated in the literature.

We will start in section 2 with the ordinary $N\!=\!1$ superstring,
for which we will be most explicit since it already displays the
relevant features. In the following section, we will then consider
the Hamiltonian reduction leading to the $N\!=\!2$ string, after
having first constructed its twisted $N\!=\!4$ symmetry algebra (such
an algebra was constructed so far only for the critical $N\!=\!2$
string \cite{suz,giro}). In section 4 we discuss the generalization
to arbitrary $N$-extended superstrings, and in section 5 we will
consider topological strings. Even though we find an analogous, but
now doubly twisted $N\!=\!4$ superconformal symmetry in these
theories, we did not quite succeed to obtain the relevant realization
of this algebra from Hamiltonian reduction.

%
%
\section{N=1 Superstring}

The idea of quantum Hamiltonian reduction is to first choose an
embedding of $s\ell(2)$ into a given Lie (super-)algebra and an
element
of the Cartan sub-algebra (that necessarily contains the Cartan
generator of the $s\ell(2)$). This Cartan element will give rise to a
splitting of the Lie algebra into sub-algebras with positive, zero
and
negative grades. Imposing first-class constraints on the
negative-grade part induces a gauge symmetry of the associated
WZW-model. In particular, one always constrains the lowering operator
of the $s\ell(2)$ to be a constant, in order to ensure the existence
of
a Virasoro generator. The other non-trivial elements of the BRST
cohomology of this gauged WZW-model then form an extension of the
Virasoro algebra \cite{hamr}.

Before applying these ideas to $osp(3|2)$, we first give a brief
description of its current algebra (the general conventions for
$osp(N|2)$ current algebras are summarized in appendix A). The
bosonic part of $osp(3|2)$ consists of an $s\ell(2)$ current algebra
at
level $\kappa$ and an $so(3)$ current algebra at level $-2\kappa$. It
is convenient for our purposes to choose a basis of the $so(3)$
sub-algebra by $J^{\pm}=(J^1\pm i J^2)$, $J^0 = i J^3$. The fermionic
part transforms according to the spin-$\frac{1}{2}$ representation of
$s\ell(2)$ and as a vector under $so(3)$. We will denote the
fermionic
currents by $j^{ab}$ with $a\in\{+,-\}$ and $b\in\{+,0,-\}$. The
generators of the Lie superalgebra are denoted by $e_\pm,e_0$,
$t^J_\pm,t^J_0$ and $t_{ab}$.

A natural choice for the $s\ell(2)$ embedding is to take just the
bosonic $s\ell(2)$ sub-algebra of $osp(3|2)$. It was shown previously
\cite{N4hamr} that such a quantum Hamiltonian reduction of $osp(N|2)$
gives rise to the standard $N$-extended superconformal algebra, if
one uses the gradation corresponding to the Cartan element $E^0$.
Taking over the general line of argumentation of \cite{blnw}, we
expect that different choices of gradations will result in different
realizations of the same extended Virasoro algebra. Indeed we will
find that if we choose instead the Cartan element $E^0+J^0$ for the
gradation, we precisely obtain the non-standard realization of the
$N\!=\!3$ superconformal algebra that represents the BRST algebra of
the $N\!=\!1$ superstring.

There are, however, differences and new features as compared to the
$s\ell(n|n\!-\!1)$-reductions, which describe the $W_n$ strings.
First, in \cite{blnw,ki} the possibility of having various, genuinely
different free-field realizations of the same superconformal algebra
was attributed to the possibility of gauging different Borel
subgroups, which for $s\ell(n|n\!-\!1)$ happens to be equivalent to
choosing different gradations. However, in the present context of
$osp(N|2)$, it appears that the realizations we are after cannot be
obtained by simply gauging different Borel subgroups. In other words,
we find that choosing different gradations is a more general method
than gauging different Borel sub-algebras.

In addition, the superconformal algebras that one obtains by
Hamiltonian reduction will always be non-linearly \cite{N4nonlin}
generated, whereas the twisted $N\!=\!3$ algebra of the $N\!=\!1$
string
(and similarly the twisted $N\!=\!4$ algebra of the $N\!=\!2$ string
discussed in chapter $3$) is linear. This is because the supercharges
that represent the BRST-current and the anti-ghost are supposed to be
nilpotent, and this is only the case for the linear form of the
$N$-extended superconformal algebras. Moreover, the numbers of the
free fermions that we get from the reduction will a priori be less
than
the number of the fields of the string model. Therefore we have to
adjoin additional fermions in a rather special way to linearize the
algebra and to obtain the proper free field realization that can be
attributed to the string theory; this is essentially the inverse of
the fermion-decoupling procedure of \cite{gosw}. We will expand
on these points later on.

Adopting the Cartan element $E^0+J^0$, we have the
following grades of the currents:\\[1cm]
\begin{tabular}{|c||c|c|c||c|c|c||c|c|c|c|c|c|}\hline
$
$&$E^+$&$E^0$&$E^-$&$J^+$&$J^0$&$J^-$&$j^{++}$&
$j^{+0}$&$j^{+-}$&$j^{-+}$&$
j^{-0}$&$j^{--}$\\ \hline
$E^0$&$1$&$0$&$-1$&$0$&$0$&$0$&$1/2$&
$1/2$&$1/2$&
$-1/2$&$-1/2$&$-1/2$\\ \hline
$J^0$&$0$&$0$&$0$&$1$&$0$&$-1$
&$1$&$0$&
$-1$&$1$&$0$&$-1$\\ \hline
$E^0+J^0$&$1$&$0$&$-1$&$1$&$0$&$-1$&$3/2$&$1/2$&$-1/2$&
$1/2$&$-1/2$&$-3/2$\\ \hline
\end{tabular}\\[1cm]
A consistent set of first-class constraints is given by
$\Phi^{\alpha}=0$,
where
\bea
\Phi^-_E &=& E^- - \frac{\kappa}{2},\non\\
\Phi^-_J &=& J^- - \la,\non\\
\Phi^{+-} &=& j^{+-} - \psi,\non\\
\Phi^{-0} &=& j^{-0} - \tau - \psib\la,\non\\
\Phi^{--} &=& j^{--}\ .
\eeal{cnstr32}
The OPE's for the auxiliary fields are
\be
\psi(z)\psib(w) = -\frac{1}{(z-w)}\,,\ \tau(z)\tau(w) =
-\frac{1}{8}\frac{\kappa}{(z-w)}\,,\
\la(z)\lab(w) =  \frac{1}{(z-w)}\,.
\eel{auxope}
The auxiliary field $\tau$ is necessary in order to make the
constraints first-class. On the other hand, the origin of the other
auxiliary fields $\psi$, $\psib$, $\la$, $\lab$ is quite different
(the latter does not even appear in the constraints). To see this,
note that \equ{cnstr32} would still form a closed set
of first class constraints even when setting $\psi$, $\psib$, and
$\la$ to zero. However, this would rather correspond to a different
$s\ell(2)$ embedding and thus would lead, as a consequence, to a
different extended superconformal algebra.

But we really want the embedding being given by $E^+\,,E^0\,,E^-$.
{}From general theorems about quantum Hamiltonian reduction, we know
that to each highest weight of the embedded $s\ell(2)$, there
corresponds a generator of the extended Virasoro algebra. Hence we
cannot constrain $J^-$ and $j^{+-}$ to be zero and simultaneously
have that these fields survive in the highest weight gauge. The
resolution of this problem becomes more clear by writing down a
Lagrangian for the constrained WZW-model of $osp(3|2)$:
\be
S= S_{WZW}[g] + \frac{1}{2\pi}\int d^2z\, str(\bar{A}\Phi) -
\frac{1}{2\pi}\int d^2z\, (\lab\delb\la + \psib\delb\psi) -
\frac{2}{\kappa\pi}\int d^2z\,
\tau\delb\tau
\eel{action}
It is gauge invariant provided the fields transform according to
\bea
\delta \tau = \frac{\ka}{2}\eta^{+0}\,,\non\\
\delta \psi = \eta^{+0}\la\,,\qq \delta \psib = 2\eta^{+-}\,,\non\\
\delta \la = 0\,, \qq \delta \lab = 2\eta^{+}_J +
\eta^{+0}\psib\,,\non\\
\delta \bar{A} = \delb \eta + [\bar{A},\eta]\,,
\qq \delta J = \del J + [J,\eta]\,.
\eeal{gaugetr}
Here, $\bar{A}$ is the gauge field that arises as Lagrange multiplier
for imposing the constraints, $J=\frac{\ka}{2} \del g g^{-1}$ denotes
the
currents of the WZW-model and $\eta$ is the gauge parameter which
takes values only in the positively graded part of $osp(3|2)$.
Several features of this lagrangian are remarkable. First we see that
despite we do not need the field $\lab$ for writing down the
constraints it is nessecary for having gauge invariance. Note also
that $\psi$, which corresponds to the the highest weight $j^{+-}$,
transforms non-trivially. Nevertheless, in the highest weight gauge
we
have $\eta^{+0}=-\frac{2}{\ka}\tau$ and therefore $\delta\psi = -
\frac{2}{\ka}\tau\la$, which
shows that $\psi$ is indeed non-zero and corresponds to a generator
of the $N\!=\!3$ algebra.

To quantize the action \equ{action},
we use the BRST-formalism and introduce ghosts and anti-ghosts as
follows:
\bea
C = ce_+ + c^J t^J_+ + \gamma^{-+}t_{-+} + \gamma^{+0} t_{+0} +
\gamma^{++} t_{++}\,,\non\\
B = b e_- + b^J t^J_- + \beta^{+-}t_{+-} + \beta^{-0} t_{-0} +
\beta^{--} t_{--}\,,
\eeal{ghosts}
The BRST-transformations are:
\bea
sB = D &,& s\bar{A} = - \delb C - [\bar{A},C] \,,
\eea
We impose the light-cone gauge by adding to \equ{action}
\be
S_{g.f.} = s \frac{1}{2\pi}\int d^2z\, B\bar{A} = \frac{1}{2\pi}\int
d^2z\, (D\bar{A} + B\delb C + \bar{A}\{B,C\}) \,.
\eel{Sgf}
{}From this we can read off the OPE's for the ghosts
\bea
b(z)c(w) = -b^J(z) c^J(w) = - \beta^{+-}(z)\gamma^{-+}(w) =
\beta^{--}(z)\gamma^{++}(w) = \,,\non\\
\qquad\qquad\qquad\qquad 2 \beta^{-0}(z)\gamma^{+0}(w) =
\frac{1}{(z-w)}\,.
\eea
Integrating out the gauge field then yields
\be
D = sB = \hat{\Phi}
\eel{bconstr}
where $\hat{\Phi}$ denotes the substitution of $J$ by $\hat{J} = J +
J_{gh}$ and
$J_{gh}=\frac{1}{2} \{B,C\}$\,. Explicitly, the ghost currents are
\bea
E^+_{gh} &=& - \frac{1}{2} \beta^{+-}\gamma^{++}\,,\non\\
E^0_{gh} &=& - \frac{1}{2} bc + \frac{1}{4}\beta^{+-}\gamma^{-+} +
\frac{1}{2}\beta^{-0}\gamma^{+0} + \frac{1}{4}\beta^{--}\gamma^{++}
\,,\non\\
E^-_{gh} &=& \frac{1}{2} \beta^{--}\gamma^{-+}\,,\non\\
J^+_{gh} &=& - \frac{1}{2} \beta^{-0}\gamma^{++}\,,\non\\
J^0_{gh} &=& - \frac{1}{2} b^Jc^J - \frac{1}{4}\beta^{+-}\gamma^{-+}
+ \frac{1}{2}\beta^{-0}\gamma^{+0} \,,\non\\
J^-_{gh} &=& - \frac{1}{2} \beta^{--}\gamma^{+0}\non\\
j^{+0}_{gh} &=& \frac{1}{4} b^J\gamma^{++} - \frac{1}{4}\beta^{+-}c^J
+ \frac{1}{2}\beta^{-0}c\,,\\
j^{+-}_{gh} &=& -\frac{1}{2} b^J\gamma^{+0} +
\frac{1}{2}\beta^{--}c\,,\non\\
j^{-+}_{gh} &=& \frac{1}{2} b\gamma^{++} +
\frac{1}{2}\beta^{-0}c^J\,,\non\\
j^{-0}_{gh} &=& \frac{1}{4} b\gamma^{+0} + \frac{1}{4}b^J\gamma^{-+}
- \frac{1}{4}\beta^{--}c^J\,,\non
\eeal{ghostcur}
Eq.\ \equ{bconstr} shows that the anti-ghosts and the constraint
currents form BRST doublets and thus decouple from the BRST
cohomology. The ghost contributions to the total currents $\hat{J}$
modify the central extensions of the algebra. In particular, we have
the following modified central terms for the Cartan currents:
\bea
\hat{E}^0(z)\hat{E}^0(w) =
\frac{(1+2\ka)}{16}\frac{1}{(z-w)^2}\,,\non\\
\hat{J}^0(z)\hat{J}^0(w) = -\frac{(1+2\ka)}{16}\frac{1}{(z-w)^2}\,,
\eea
It is now straightforward to write down the BRST operator:
\bea
{\cal Q}_{BRST}&=& \frac{1}{2\pi i} \oint d^2z\,\Big[c(E^- +
\frac{1}{2}E^-_{gh} -\frac{\kappa}{2}) -c^J(J^- +
\frac{1}{2}J^-_{gh}-\la) +
\gamma^{-+}(j^{+-} + \non\\
& &+ \frac{1}{2}j^{+-}_{gh}-\psi) -
2\gamma^{+0}(j^{-0} + \frac{1}{2}j^{-0}_{gh}-\tau -
\frac{1}{4}\psib\la) - \gamma^{++}j^{--}\,\Big]\,,
\eeal{brstop}
which can be split it into the three following pieces:
\bea
{\cal Q}_0 &=& \frac{1}{2\pi i} \oint d^2z\, (-\frac{\kappa}{2} c +
c^J \la)\,,\non\\
{\cal Q}_1 &=& \frac{1}{2\pi i} \oint d^2z\, (-\gamma^{-+}\psi + 2
\gamma^{+0}\tau +
\frac{1}{2} 2 \gamma^{+0}\psib\la)\,,\non\\
{\cal Q}_2 &=& \frac{1}{2\pi i} \oint d^2z\,\Big[c(E^- +
\frac{1}{2}E^-_{gh}) -c^J (J^-+   \frac{1}{2}J^-_{gh}) +
\gamma^{-+}(j^{+-} + \frac{1}{2}j^{+-}_{gh}) -\non\\ &
&2\gamma^{+0}(j^{-0} + \frac{1}{2}j^{-0}_{gh}) -
\gamma^{++}j^{--}\,\Big]\,,
\eeal{filtration}
where ${\cal Q}_0^2 = {\cal Q}_2^2 =0$ and ${\cal Q}_1^2 + \{{\cal
Q}_0,{\cal Q}_2\} = 0$. This shows
that the techniques developed in \cite{hamr} can be fully taken over.
In particular, the spectral sequence techniques apply and we can
define the quantum Miura transformation as the truncation of the
generators of the cohomology of \equ{brstop} to the zero-grade
fields,
which are given by the the Cartan generators of $osp(3|2)$ and the
auxiliary fields. This method guarantees that we end up with a
free-field realization of the non-linear $N\!=\!3$ superconformal
algebra. Note that this is specifically due to our gradation
$E^0+J^0$ (for the usual gradation $E^0$ \cite{N4hamr}, the whole
$so(N)$ sub-algebra of $osp(N|2)$ has grade zero and thus does not
get
bosonized).

Using the methods of \cite{hamr}, where we use the filtration of
${\cal Q}_{BRST}$ given by \equ{filtration}, we can compute the
non-trivial elements of the cohomology by starting with the total
currents that are $s\ell(2)$ highest weights. In this way, we find
for the stress tensor:
\bea
T &=& \frac{4\ka}{1+2\ka}\, \Big( \hat{E}^+  +
\frac{2}{\ka}\hat{E}^{0^2}- \del \hat{E}^0 -
\frac{2}{\ka}\hat{J}^{0^2} + \alpha \del \hat{J}^0 -
\frac{1-2\alpha}{8\ka}\del\la\lab + \non\\ & & +
\frac{1+4\ka+2\alpha\ka}{8\ka}\la\del\lab +
\frac{1+2\ka}{k^2}\tau\del\tau +  \frac{1+3\ka+\alpha\ka}{4\ka}
\psi\del\psib +\frac{1+\alpha}{4}\del\psi\psib -\non\\& &-
\frac{4}{\ka} \hat{j}^{+0}\tau + \frac{2}{\ka}\hat{j}^{-+}\psi -
\frac{2}{\ka}\hat{J}^+\la + \frac{1}{\ka}\hat{j}^{+0}\la\psi\Big)
\eeal{TnlN3}
and for the generators of the $so(3)$ Kac-Moody algebra:
\bea
K^+ &=& 4 \Big(\hat{J} + \frac{2}{\ka}\hat{j}^{-+}\tau -
\frac{1}{2\ka}\hat{j}^{-+}\la\psib - \frac{1}{\ka}\hat{E}^0\psib\tau
- \frac{1}{2}\hat{J}^0\lab -
\frac{1}{2}\hat{j}^{+0}\psib - \frac{1}{\ka}\hat{J}^0\psib\tau
+\non\\ & & +\frac{1}{2\ka}\hat{j}^{-+}\psib\la -
\frac{1+2\ka}{8}\del\lab - \frac{1+2\ka}{4}\del\psib\tau -
\frac{1}{16}\lab^2\la - \frac{1}{8}\lab\psi\psib\Big)\,,\\
K^3 &=& 4(\hat{J}^0 + \frac{1}{4}\lab\la - \frac{1}{4}\psib\psi)
\,,\\
K^- &=& 4 \la\ .
\eeal{KnlN3}
In addition, we get the following three supercurrents:
\bea
G^+ &=& \frac{4 i \sqrt{2}}{\sqrt{1+2\ka}} \Big( \hat{j}^{++} -
\frac{(1+2\ka)^2}{16\ka}\del^2\psib + \frac{1}{2}\hat{E}^+\psib -
\frac{2}{\ka}\hat{J}^+\tau - \frac{1}{2}\hat{j}^{+0}\lab - \non\\
& & - \hat{E}^0\hat{j}^{-+} +
\frac{2}{\ka}\hat{J}^0\hat{j}^{-+} + \frac{1}{\ka}\hat{E}^{0^2}\psib
+ \frac{1}{\ka}\hat{J}^{0^2}\psib - \frac{1}{2}\del{E}^0\psib -
\frac{1}{2}\del\hat{J}^0\psib +\non\\
& & + \frac{1+2\ka}{8\ka}\psi\del\psib\psib
+ \frac{1}{\ka}\hat{j}^{+0}\psib\tau -
\frac{1}{\ka}\hat{j}^{-+}\psib\psi +
\frac{1}{2\ka}\hat{j}^{+0}\lab\la -\non\\
& & -\frac{1+2\ka}{16\ka}\del\psib\lab\la -
\frac{1}{16}\la^2\psi + \frac{1}{\ka}\hat{E}^0\lab\tau -
\frac{1}{4\ka}\hat{J}^0\psib\lab\la - \non\\
& & - \frac{1+2\ka}{4\ka}\lab\del\tau -
\frac{1}{4\ka}\hat{E}^0\psib\la\lab - \frac{1}{8\ka}\lab^2\la\tau +
\frac{1}{4\ka}\psib\psi\lab\tau - \frac{1}{4\ka}\del\psib\hat{E}^0 -
\non\\
& & -\frac{3+4\ka}{4\ka}\psib\hat{J}^0 + \del \hat{j}^{-+}\Big)\,,\\
G^3 &=& \frac{4 i \sqrt{2}}{\sqrt{1+2\ka}} \Big(\hat{j}^{+0}  -
\frac{1}{\ka}\hat{j}^{-+}\la + \frac{1+2\ka}{8\ka}\del\psib\la +
\frac{1}{4}\lab\psi - \frac{2}{\ka}\hat{E}^0\tau +
\frac{1}{2\ka}\hat{J}^0\psib\la +\non\\
& &+\frac{1+2\ka}{2\ka}\del\tau + \frac{1}{2\ka}\hat{E}^0\psib\la +
\frac{1}{2\ka}\lab\la\tau - \frac{1}{2\ka}\psib\psi\tau\Big)\,,\\
G^- &=& \frac{4 i \sqrt{2}}{\sqrt{1+2\ka}}(\psi +
\frac{2}{\ka}\tau\la)\,.
\eeal{GnlN3}
{}From these expressions we get a free-field realization of the
$N\!=\!3$ superconformal algebra in nonlinear form, by taking only
the zero grade part of each generator. However, a simple count of the
number of fields shows that there is one fermion missing to give the
field content of the $N\!=\!1$ string (assuming that the matter
sector is realized in terms of one boson and one fermion). Of course,
this is not surprising since the non-linear form of the algebra is
well-known to arise by factorization of a single fermion \cite{gosw}.

Note also that in the expression for the stress tensor there is a
free parameter $\alpha$, which determines the conformal weights of
$K^+$, $K^-$, $G^+$, $G^-$, and thus serves as a twist parameter.
If we choose the value $\alpha =1+\frac{1}{\ka}$, we get
conformal weight $1/2$ for $K^+$, $3/2$ for $K^-$, $1$ for $G^+$ and
$2$ for $G^-$. This choice is motivated by the string representation,
where $G^-$ is the anti-ghost. The resulting central
charge of the stress tensor is then $-1/2$. Since in a string theory
the central extension of the Virasoro algebra should vanish, this is
further indication of the missing of one fermionic degree of freedom.
Therefore, we adjoin a free fermion $\rho$ with OPE
\be
\rho(z)\, \rho(w) = -\frac{(1+2\ka)}{(z-w)}\,.
\ee
We can now linearize the algebra by redefining the generators
\bea
T_{lin} &=& T + \frac{1}{2(1+2\ka)}\rho\del\rho \,,\non\\
G^a_{lin} &=& G^a + \frac{1}{1+\ka}\rho K^a\,.
\eea
This indeed results in a free-field realization of the linear
$N\!=\!3$ algebra (cf., (B.2)). However, the generator $G^-$ is not
yet precisely of the desired form, for which $G^- \propto \psi$.
Rather, we obtain the expression
\be
G^- = \frac{4 i \sqrt{2\ka}}{\sqrt{1+2\ka}}(\psi +
\frac{2}{\ka}\tau\la) - \frac{4}{1+2\ka} \rho\la\,.
\ee
{}From this we can infer that we need, in addition, to perform a
similarity transformation
\bea
\tilde{T} &=& S T S^{-1}\,,\non\\
\tilde{G^a} &=& S G^a S^{-1}\,,\non\\
\tilde{K^a} &=& S K^a S^{-1}\,,\non\\
F &=& S \rho S^{-1}\,,
\eea
where
\be
S = \exp\Big[\frac{1}{2\pi i} \oint dz\,
\big(\frac{2}{\ka}\tau\la\psib -
\frac{i}{\sqrt{2\ka(1+2\ka)}}\rho\la\psib\big)\Big]\  .
\ee
Furthermore, we bosonize the Cartan currents as follows:
\bea
\hat{E}^0 = i \frac{\sqrt{1+2\ka}}{4} \del \varphi_1 &,&
\hat{J}^0 = \frac{\sqrt{1+2\ka}}{4} \del \varphi_2\,,\non\\
\varphi_k(z)\, \varphi_l(w) = -\delta_{kl} \ln (z-w)\,,
\eea
and rescale the remaining fields according to
\bea
\rho \rightarrow \frac{1}{\sqrt{1+2\ka}}\,\rho &,&\tau \rightarrow
\frac{2\sqrt{2}}{\sqrt{\ka}}\, \tau \,,\non\\
\psi \rightarrow \frac{2\sqrt{2\ka}}{\sqrt{1+2\ka}}\, \psi &,&
\psib \rightarrow -\frac{\sqrt{1+2\ka}}{2\sqrt{2\ka}}
\,\psib\,,\non\\
\la \rightarrow -2 i\la &,& \lab \rightarrow \frac{i}{2} \lab\,.
\eea
Then, after all these manipulations, we finally arrive at the
following form for the Miura-transformed linearized generators:
\bea
\tilde{T} &=& -\frac{1}{2}(\del\varphi_1)^2
-\frac{1}{2}(\del\varphi_2)^2 + \frac{3}{2}\la\del\lab +
\frac{1}{2}\del\la\lab - 2 \psi\del \psib - \del \psi \psib +\non\\ &
& +\frac{1}{2}\rho\del\rho +\frac{1}{2}\tau\del\tau -
\frac{i\ka}{\sqrt{1+2k}}\del^2\varphi_1 +
\frac{1+\ka}{\sqrt{1+2k}}\del^2\varphi_2 \,,\non\\
\tilde{G}^+ &=& 2i \Big( \psib (\frac{1}{2}(\del\varphi_1)^2 +
\frac{1}{2}(\del\varphi_1)^2 +
\frac{i\ka}{\sqrt{1+2\ka}}\del^2\varphi_1 -
\frac{1+\ka}{\sqrt{1+2k}}\del^2\varphi_2 -
\frac{3}{2}\la\del\lab - \non\\
& &-\frac{1}{2}\del\la\lab -
\frac{1}{2}\rho\del\rho -\frac{1}{2}\tau\del\tau ) + \lab
(\frac{1}{2}\del\varphi_1\tau +
\frac{1}{2}\del\varphi_2\rho +
\frac{i\ka}{\sqrt{1+2k}}\del\tau -\non\\
& & -\frac{1+\ka}{\sqrt{1+2\ka}} \del \rho) -
\frac{1}{2}(\frac{3}{2}\psib\la\del\lab +
\frac{1}{2}\psib\del\la\lab - \psib\psi\del\psib +
\frac{3}{2}\lab\la\del\psib + \non\\
& & + \frac{1}{2} \lab^2\psi + \psib\lab\del\la) +
\frac{\sqrt{1+2k}}{2}\del(\lab\rho) + \frac{1+2k}{2} \del^2\psib +
\frac{1}{4}\del(\lab\la\psib) + \non\\
& & + \sqrt{1+2\ka} \del(\varphi_2\psib)\Big)\,,\non\\
\tilde{G}^3 &=& \del\varphi_1\tau + \del\varphi_2\rho + 3
\la\del\psib + 2 \del\la\psib +\psi\lab + \frac{2i
\ka}{\sqrt{1+2\ka}} \del \tau - \frac{2+2\ka}{\sqrt{1+2\ka}} \del
\rho\,,\non\\
\tilde{G}^- &=& 2i \psi\,,\non\\
\tilde{K}^+ &=& 2i \Big(\!-\frac{1}{2}\psib\del\varphi_1\tau
-\frac{1}{2}\psib\del\varphi_2\rho +\frac{1}{2}\psib\la\del\psib +
\frac{1}{2}\psib\lab + \frac{i\ka}{\sqrt{1+2\ka}}\psib\del\varphi_1 +
\non\\
& & +\frac{\ka}{\sqrt{1+2\ka}}\psib\del\varphi_2 -
\frac{\sqrt{1+2\ka}}{2}\del\varphi_2\lab + \frac{1}{4}\lab^2\la  +
\frac{\sqrt{1+2\ka}}{2}\del\psib\rho + \non\\
& &+\frac{1+2\ka}{2}\del\lab\Big)\,,\non\\
\tilde{K}^3 &=& - \psi\psib + \la\lab +
\sqrt{1+2\ka}\del\varphi_2\,,\non\\
\tilde{K}^- &=& 2i \la\,,\non\\
F &=& \la\psib + \sqrt{1+2\ka}\rho\,.
\eea
These formulas correspond exactly to the generators of the hidden
$N\!=\!3$ superconformal algebra of the $N\!=\!1$ superstring as
found in \cite{blnw}. We just need to identify $\varphi_1$ and $\tau$
with the matter system, and $\varphi_2$ and $\rho$ with the Liouville
system. Moreover, we see that the diffeomorphism ghost of the
superstring is represented by $\psib$, the anti-ghost by $\psi$, and
their bosonic superpartners by $\lab$ and $\la$, respectively. This
means that $G^+$ is indeed the $N\!=\!1$ string BRST operator,
$K^+$ its superpartner with respect to the $N\!=\!1$ supercharge,
$G^3$, and $K^3$ is the ghost number current. Finally, note that the
Kac-Moody level $\ka$ is related to the matter central charge as
follows:
\be
\hat c_m\ =\ 5-2\big((1+2\ka)+\frac{1}{1+2\ka}\big)\ ,
\ee
where, of course, $\hat c_m=10$ corresponds to the critical
superstring.
%
%
\section{N=2 Superstring}
In this section we will generalize the methods discussed above to the
$N\!=\!2$ superstring \cite{adetal},\cite{N2ss}. We will first
construct the generators of the hidden, twisted $N\!=\!4$
superconformal algebra. In fact, such an algebra has already been
found in \cite{giro}, but only for critical $N\!=\!2$ strings. We
will generalize this to non-critical $N\!=\!2$ strings, where the
conformal anomaly is cancelled by the $N\!=\!2$ Liouville system.

To begin, we would like to make some general remarks about the
twisted $N\!=\!4$ algebra. (Our conventions, in which we follow
\cite{guen}, can be found in appendix B.) It is known \cite{setp}
that
$N\!=\!4$ algebras come in a lot of varieties, but the most general
algebra consists of the energy-momentum tensor $T$, four
supercurrents $G_\pm$, $G_{\pm K}$, and two commuting $s\ell(2)$
Kac-Moody algebras at levels $k^+$ and $k^-$ whose currents we denote
by $A^{\pm i}$ with $i \in \{+,3,-\}$. Furthermore, there are four
fermionic currents of spin $\frac{1}{2}$, denoted by $Q_\pm$, $Q_{\pm
K}$, plus a U(1) current $U$. It is possible to redefine the
generators such that the spin-$\frac{1}{2}$ currents and the U(1)
current decouple, so that one is left over with the energy momentum
tensor, the supercharges and the Kac-Moody currents. This is the
non-linear form of the algebra, which is the form that we will get
from Hamiltonian reduction of $osp(4|2)$.

Let us first discuss the twisted linear $N\!=\!4$ algebra, by noting
that there are two separate $N\!=\!2$ sub-algebras. We can use
one of the two associated $U(1)$ currents to twist the
energy-momentum tensor as follows:
\be
\Th = T + \del M_{+K-K}\,,
\ee
where the $N\!=\!4$ current $M$ is defined in appendix B.
The generators $\Th$, $G_{\pm K}$ and $2M_{+K-K}$ then form a
topologically twisted $N\!=\!2$ sub-algebra. The other $N\!=\!2$
algebra has to have vanishing central charge in order to be
consistent with the vanishing conformal anomaly of $\Th$.
We can achieve this by redefining
\be
\hat{G}_\pm = G_\pm - \frac{2k^\pm}{k}\del Q_\pm\,.
\ee
Moreover, the condition for $\hat{G}_\pm$ to be primary with respect
to $\Th$ forces the levels of the Kac-Moody algebras to be equal:
$k^+=k^-$. In addition, due to these modifications, we also have to
redefine the $U(1)$ current
\be
\hat{M}_{+-} = M_{+-} + \frac{U}{2}\,.
\ee
The generators $\Th$, $\hat{G}_\pm$ and $\hat{M}_{+-}$ form a
$N\!=\!2$ superconformal algebra with vanishing central charge. After
twisting, the conformal weight of $G_{+K}$ is equal to $1$,
$G_{-K}$ has weight $2$, $A^{++}$ and $A^{--}$ have weights
$\frac{1}{2}$, and $A^{+-}$ and $A^{-+}$ have weights $\frac{3}{2}$.
Furthermore, $Q_{+K}$ and $Q_{-K}$ have weights $0$ and $1$,
respectively. The conformal weights of the rest of the generators
remain unchanged.

For the construction of this algebra in terms of the field content of
a $N\!=\!2$ string theory we follow the line of \cite{blnw}. We do
not make any assumption about the specific realization of
the matter system. We represent it, generically, by the generators of
a $N\!=\!2$ superconformal algebra: $T_m$, $G_{m\pm}$ and $J_m$, and
denote its central charge by $c_m$.

The Liouville system is given in terms of a complex scalar field
$\fl$ and a complex spinor $\pl$ with OPE
\bea
\del\fl(z)\del\bfl(w) = -\frac{1}{(z-w)^2}
 \,,&& \pl(z)\bpl(w) = \frac{1}{z-w} .
\eea{}
In terms of these fields we can form an $N\!=\!2$ algebra as follows:
\bea
J_l &=& - \pl\bpl - Q \del\fl - Q \del\bfl ,\\
G_{l+} &=& \del\fl \bpl - Q \del \bpl ,\\
G_{l-} &=& -\del\bfl \pl - Q \del \pl ,\\
T_l &=& - \del\fl\del\bfl
-\frac{1}{2}\pl\del\bpl+\frac{1}{2}\del\pl\bpl -
\frac{Q}{2}\del^2\fl + \frac{Q}{2}\del^2\bfl ,
\eea{}
where we introduced a background charge parameter, $Q$, such that the
central charge of the Liouville system is given by $c_l = 3-6Q^2$.

Since we are dealing with local $N\!=\!2$ supersymmetry on the string
world sheet, we introduce a $(b,c)$ ghost system for the
diffeomorphism symmetry, two bosonic ghost systems
$(\beta_\pm,\gamma_\mp)$ for the super-diffeomorphisms, and
additional
fermionic ghosts $(\eta,\xi)$ with spins $(1,0)$ for the U(1) gauge
symmetry. The OPE's of these ghosts are
\be
b(z)c(w) = \bpm(z)\gamma_{\mp}(w) =\eta(z)\xi(w) = \frac{1}{z-w}\  .
\ee
We then find for the $N\!=\!2$ algebra in the ghost sector
\bea
J_g &=& -\del(\eta c) - \bp\gm + \bm\gp ,\\
G_{g+} &=& \frac{3}{2}\bp\del c + \del\bp c +\frac{1}{2}\del\eta \gp
+
\eta\del\gp + \bp\xi - b\gp ,\\
G_{g-} &=& -\frac{3}{2}\bm\del c - \del\bm c + \frac{1}{2}\del\eta
\gm +
\eta\del\gm + \bm\xi + b\gm ,\\
T_g &=& -2 b\del c - \del b c + \frac{3}{2} \bp \del\gm +
\frac{1}{2} \del\bp \gm +\frac{3}{2} \bm \del\gp +\frac{1}{2}
\del\bm\gp - \non \\& &-\eta\del\xi ,
\eea{}
which has a central charge equal to $-6$. The total central charge of
the $N\!=\!2$ string, joining together matter, Liouville fields and
ghosts, thus vanishes, provided we fix the Liouville background
charge
to be $Q= \pm\frac{\sqrt{c_m-3}} {\sqrt{6}}$. From the ghosts we can
build an additional $N\!=\!2$ multiplet of currents
\bea
\jg &=& - \eta c ,\\
\jgp &=& \eta\gp + \bp c ,\\
\jgm &=& \eta\gm - \bm c ,\\
\Jg &=& -bc +\bp\gm + \bm\gp - \eta\xi .
\eea{}
where the highest-spin component is the ghost number current.
The $N\!=\!2$ string BRST-current is part of an $N\!=\!2$ multiplet
as well, which has components:
\bea
j_{BRST} &=& c (J_m + J_l + \frac{1}{2}J_g) ,\\ j_{BRST+} &=& \gp
(J_m + J_l + \frac{1}{2}J_g) - c (G_{m+} + G_{l+} +
\frac{1}{2}G_{g+}) ,\\ j_{BRST-} &=& \gm (J_m + J_l + \frac{1}{2}J_g)
- c (G_{m-} + G_{l-} + \frac{1}{2}G_{g-}) ,\\ J_{BRST} &=& c (T_m +
T_l + \frac{1}{2}T_g) +\gp(G_{m-} + G_{l-} + \frac{1}{2}G_{g-})-\non
\\ & & -\gm(G_{m+} + G_{l+} + \frac{1}{2}G_{g+}) + \xi(J_m + J_L +
\frac{1}{2}J_g) .
\eeal{n2brst}
Let us now introduce the modified ghost current
\be
\I = \Jg - Q \del\fl + Q \del\bfl ,
\ee
and the improved BRST current
\bea
\Ibrs &=& J_{BRST} - \frac{1}{2}\del(c\xi\eta) -
\frac{1}{4}\del(c\gp\bm)
- \frac{1}{4}\del(c\gm\bp) + Q^2 \del^2 c + \non\\
 & &+ Q\del(c\del\fl) - Q\del(c\del\bfl) + Q\del(\gp\pl) -
Q\del(\gm\bpl) ,
\eea{}
where the total derivative terms have been chosen such that $\Ibrs$
is a primary field of weight $1$ and has regular OPE with itself.
Together with the total energy momentum tensor, $T_{tot}= T_m + T_l +
T_g$, and the antighost, $b$, these currents form a topologically
twisted $N\!=\!2$ algebra with central extension
\be
c = c_m -3\ .
\ee
Note that the critical $N\!=\!2$ string, which has $c_m=3$, is mapped
onto a twisted $N\!=\!4$ algebra with vanishing central charge; it is
this case which has previously been discussed in \cite{giro}.

In order to make contact with the above-described twisted $N\!=\!4$
algebra, we identify
\be
T_{tot}\leftrightarrow \hat{T} \;,\;\Ibrs \leftrightarrow
G_{+K}\;,\;b \leftrightarrow G_{-K}\;,\;
\I \leftrightarrow 2M_{+K-K} .
\ee
Furthermore, we can identify the original $N\!=\!2$ generators of the
string, $G_{tot\pm} = G_{m\pm} + G_{l\pm} + G_{g\pm}$
and $J_{tot} = J_m + J_l + J_g$, with the twisted currents of the
$N\!=\!4$ superconformal algebra:
\bea
G_{tot\pm} \leftrightarrow \Gh_\pm\,,&
J_{tot}  \leftrightarrow \hat{M}_{+-}\,.
\eea{}
Let us briefly sketch how to construct the remaining $N\!=\!4$
currents. One reads from the twisted $N\!=\!4$ superconformal algebra
that
\be
G_{+K}(z)\Gh_+(w) = \frac{i}{2}\frac{A^{++}(w)}{(z-w)^2} .
\ee
Similar relations hold for the other generators. With the help of
the above-mentioned identifications, one gets the whole twisted
$N\!=\!4$ algebra, described at the beginning of this section, in
terms of the fields of the $N\!=\!2$ string. In summary, we find the
following expressions for the $s\ell(2)$ Kac-Moody currents in the
($+$)-sector
\bea
A^{++} &=& i(j_{BRST+} + 2 Q \bpl\xi + Q \bpl\del c +
\frac{1}{2}\bm\gp^2 -
\frac{1}{2}\bp c\xi+ \non\\
& &+ \frac{1}{2}\bp\gp\gm + \frac{1}{4} \bp c \del c + 2 Q
\gp\del\bfl -
\frac{1}{2}\gp bc + \frac{1}{4} \gp c \del\eta - \non\\
& &-\gp\eta\xi + 2Q \del \bpl c + \frac{1}{2} \del\gp c \eta - 2 Q^2
\del\gp) ,\\
A^{+3} &=& -\frac{i}{2}( J_{tot} + \Jg -\frac{1}{2}\del\jg + 2 Q
\del\bfl ) ,\\
A^{+-} &=& -i \bm \ ,
\eeal{N2A+}
and the following expressions for the $s\ell(2)$ Kac-Moody currents
in
the ($-$)-sector:
\bea
A^{-+} &=& -i \bp ,\\
A^{-3} &=& -\frac{i}{2}(J_{tot} - \Jg -\frac{1}{2}\del\jg + 2 Q
\del\fl) ,\\
A^{--} &=& -i (j_{BRST-} - 2 Q \pl\xi + Q\pl\del c - \frac{1}{2}\bm
c\xi -
\frac{1}{2}\bm\gp\gm -\non \\
& &-\frac{1}{4}\bm c\del c - \frac{1}{2}\bp\gm^2 + 2 Q \gm\del\fl +
\frac{1}{2}
\gm bc + \frac{1}{4} \gm c\del\eta +\non \\
& & + \gm\eta\xi + 2 Q \del\pl c +
 \frac{1}{2}\del\gm c \eta + 2 Q^2\del\gm) .
\eeal{N2A-}
Furthermore, we have the four fermionic spin-$\frac{1}{2}$ currents
\bea
Q_{+K} &=& -\frac{1}{2} j_{BRST} - 2 Q^2 \xi - Q \bpl\gm - Q \pl\gp -
\frac{1}{4}\bm\gp c + \non \\
& & + \frac{1}{4} \bp\gm c - Q c \del\fl -  Q c \del\bfl +
\frac{1}{2} \gp\gm\eta +
\frac{1}{4} c\del c \eta ,\\
Q_{-K} &=& \frac{1}{2} \eta ,\\
Q_+ &=& - \frac{1}{2} \jgp + Q \bpl ,\\
Q_- &=& + \frac{1}{2} \jgm - Q \pl .
\eeal{N2ferm}
and, finally, a spin-$0$ bosonic current\be
\varphi =  \frac{1}{2} \jg - Q \fl - Q \bfl ,
\eel{N2u1}
where the $U(1)$ current of the twisted $N\!=\!4$ superconformal
algebra is given by $U = \del\varphi$.\\

In the remainder of this section we will show how one can construct
these currents by means of quantum Hamiltonian reduction of
$osp(4|2)$. The bosonic part of this Lie superalgebra consists of an
$s\ell(2)$ algebra at level $\ka$ and an $so(4)$ algebra at level $-2
\ka$. We make use of the fact that $so(4)$ can be written in terms of
two commuting $s\ell(2)$ algebras, in redefining the currents
\bea
M^\pm &=& i (J^{12} + J^{34} \pm i J^{13} \mp i J^{23})\,,\non\\
M^0 &=& i (J^{14} + J^{23})\,,\non\\
N^\pm &=& i (-J^{12} + J^{34} \mp i J^{13} \pm i J^{23})\,,\non\\
N^0 &=& -i (J^{14} - J^{23})\,.
\eeal{sl2MN}
For the fermionic currents of $osp(4|2)$, we choose a new basis as
well,
\bea
j^{\pm ab} = \sigma_i^{ab} j^{\pm i}  &,& a,b \in \{+,-\} \,,
\eea
where
\bea
\sigma_1 = \left(\begin{array}{cc}1&0\\0&1 \end{array}\right),\qquad
\sigma_2 = \left(\begin{array}{cc}0&1\\1&0 \end{array}\right),\qquad
\sigma_3 = \left(\begin{array}{cc}0&-i\\i&0 \end{array}\right),\qquad
\sigma_4 = \left(\begin{array}{cc}1&0\\0&-1 \end{array}\right)
\eea
These fermionic currents transform under the representation
$(\frac{1}{2},\frac{1}{2},\frac{1}{2})$ of the three $s\ell(2)$
sub-algebras of $osp(4|2)$.

We now define the gradation by choosing the
Cartan element $H=E^0+M^0-N^0$. Accordingly,
the algebra $osp(4|2)$ decomposes into positive, zero and
negative parts, which are displayed in the following table:

\begin{tabular}{|c||c|c|c||c|c|c||c|c|c|}\hline
$ $&$E^+$&$E^0$&$E^-$&$M^+$&$M^0$&$M^-$&$N^+$&$N^0$&$N^-$\\ \hline
$H$&$1$&$0$&$-1$&$1$&$0$&$-1$&$1$&
$0$&$-1$\\ \hline
\end{tabular}\\[0.3cm]
\begin{tabular}{|c||c|c|c|c|c|c|c|c|}\hline
$
$&$j^{+++}$&$j^{++-}$&$j^{+-+}$&$j^{+--}$&
$j^{-++}$&$j^{-+-}$&$j^{--+}$&$j^{---}$\\ \hline
$H$&$1/2$&$3/2$&$-1/2$&$1/2$&$-1/2$&$1/2$&$-3/2$&
$1/2$\\ \hline
\end{tabular}\\[1cm]
The constraints that we choose are compatible with the gradation and
are given by:
\bea
\Phi^-_E = E^- - \frac{\ka}{2}\,,\non\\
\Phi^-_M = M^- - \mu\,,\non\\
\Phi^+_N = N^+ - \nu\,,\non\\
\Phi^{+-+} = j^{+-+} - \psi\,,\non\\
\Phi^{-++} = j^{-++} - \chi - \frac{1}{4}\psib\nu\,,\non\\
\Phi^{--+} = j^{-++}\,,\non\\
\Phi^{---} = j^{---} - \chib - \frac{1}{4}\psib\mu\,,
\eeal{constrN4}
where
\bea
\chi(z)\chib(w) = -\frac{1}{4}\frac{\ka}{(z-w)}\ ,\ \ \
\psi(z)\psib(w)=\mu(z)\mub(w)=\nu(z)\nub(w)=\frac{1}{(z-w)}\ .
\eeal{constrN4ope}
As for the $N\!=\!1$ string, the $s\ell(2)$ embedding is given by
$E^\pm,E^0$, and we again face the problem of having highest weights
in the negatively graded part. In the above choice of constraints, we
took care of this by the introduction of the auxiliary fields
$\mu,\mub,\nu,\nub,\psi,\psib$; in addition, the auxiliary
fields $(\chi,\chib)$ are needed to make the constraints first class.
The corresponding gauge invariant action has the form
\be
S= S_{WZW}[g] + \frac{1}{2\pi}\int d^2z\, str(\bar{A}\Phi) -
\frac{1}{2\pi}\int d^2z\, (\mub\delb\mu + \nub\delb\nu +
\psib\delb\psi) -
\frac{2}{\kappa\pi}\int d^2z\,
\chib\delb\chi
\eel{actionN4}
Accordingly, the gauge transformation for the fields are
\beac
\delta \chi = \frac{\ka}{2}\eta^{+++}\,,\delta \chib =
\frac{\ka}{2}\eta^{+--}\,,\non\\
\delta \psi =
-\frac{1}{2}\eta^{+++}\mu-\frac{1}{2}\eta^{+--}\nu\,,\qq \delta \psib
= 2\eta^{-+-}\,,\\
\delta \mu = 0\,, \qq \delta \mub = -\eta^{+}_M +
\frac{1}{2}\eta^{+++}\psi\,,\non\\
\delta \nu = 0\,, \qq \delta \nub = -\eta^{-}_N +
\frac{1}{2}\eta^{+++}\psi\,,\non\\
\delta \bar{A} = \delb \eta + [\bar{A},\eta]\,,
\qq \delta J = \del\eta + [J,\eta]\,.
\eeacn{gaugetrN4}
We do the gauge fixing analogous to the previous chapter. In order to
obtain the BRST operator, we need the following ghosts (these are not
to be confused with the ghosts of the $N\!=\!2$ system !):
\bea
C = ce_+ + c^M t^M_+ + c^N t^N_- + \gamma^{-+-}t_{-+-} + \gamma^{+--}
t_{+--} + \gamma^{++-} t_{++-} + \gamma^{+++} t_{+++}\,,\non\\
B = b e_- + b^M t^M_- + b^N t^N_+ + \beta^{+-+}t_{+-+} + \beta^{-++}
t_{-++} + \beta^{--+} t_{--+} + \beta^{---} t_{---}\,.
\eeal{ghostsN4}
{}From the ghost action, which is the same as \equ{Sgf}, we can read
off the OPE's for the ghosts:
\bea
b(z)c(w) = \frac{1}{(z-w)}\,,\qquad b^M(z)c^M(w) = b^N(z)c^N
(w)=\frac{-2}{(z-w)}\,,\non\\ \beta^{+-+}(z)\gamma^{-+-}(w) =
\beta^{-++}(z)\gamma^{+--}(w) = \beta^{---}(z)\gamma^{+++}(w)= \non\\
\qquad\qquad\qquad\qquad\qquad= -
\beta^{--+}(z)\gamma^{++-}(w) = \frac{1}{(z-w)}\,.
\eea
The currents $J$ of $osp(4|2)$ get modified by the ghost
contributions:
$J\rightarrow\hat{J}\equiv J+J_{gh}$, where
\bea
E^+_{gh} &=& \frac{1}{2} \beta^{+-+}\gamma^{++-}\,,\non\\
E^0_{gh} &=& -\frac{1}{2} bc - \frac{1}{4}\beta^{+-+}\gamma^{-+-} +
\frac{1}{4}\beta^{-++}\gamma^{+--} -
\frac{1}{4}\beta^{--+}\gamma^{++-} +
\frac{1}{4}\beta^{---}\gamma^{+++}\,,\non\\
E^-_{gh} &=& -\frac{1}{2} \beta^{--+}\gamma^{-+-}\,,\non\\
M^+_{gh} &=& \frac{1}{2} \beta^{-++}\gamma^{++-}\,,\non\\
M^0_{gh} &=& \frac{1}{4} b^Mc^M + \frac{1}{4}\beta^{+-+}\gamma^{-+-}
- \frac{1}{4}\beta^{-++}\gamma^{+--} -
\frac{1}{4}\beta^{--+}\gamma^{++-} +
\frac{1}{4}\beta^{---}\gamma^{+++}\,,\non\\
M^-_{gh} &=& -\frac{1}{2} \beta^{--+}\gamma^{+--}\,,\non\\
N^+_{gh} &=& -\frac{1}{2} \beta^{--+}\gamma^{+++}\,,\non\\
N^0_{gh} &=& -\frac{1}{4} b^Nc^N - \frac{1}{4}\beta^{+-+}\gamma^{-+-}
- \frac{1}{4}\beta^{-++}\gamma^{+--} +
\frac{1}{4}\beta^{--+}\gamma^{++-} +
\frac{1}{4}\beta^{---}\gamma^{+++}\,,\non\\
N^-_{gh} &=& -\frac{1}{2} \beta^{--+}\gamma^{-+-}\,,\non\\
j^{+++}_{gh} &=& -\frac{1}{4} b^N\gamma^{++-} -
\frac{1}{4}\beta^{+-+} c^M + \frac{1}{2} \beta^{-++} c\,, \non\\
j^{+-+}_{gh} &=& -\frac{1}{4} b^M\gamma^{+++} - \frac{1}{4}b^N
\gamma^{+--} + \frac{1}{2} \beta^{--+} c\,, \non\\
j^{+--}_{gh} &=& -\frac{1}{4} b^M\gamma^{++-} -
\frac{1}{4}\beta^{+-+} c^N + \frac{1}{2} \beta^{---} c\,, \non\\
j^{-++}_{gh} &=& \frac{1}{2} b \gamma^{+++} - \frac{1}{4}b^N
\gamma^{-+-} - \frac{1}{4} \beta^{--+} c^M\,, \non\\
j^{-+-}_{gh} &=& \frac{1}{2} b \gamma^{++-} - \frac{1}{4}\beta^{-++}
c^N - \frac{1}{4} \beta^{---} c^M\,, \non\\
j^{---}_{gh} &=& \frac{1}{2} b \gamma^{+--} - \frac{1}{4}b^M
\gamma^{-+-} - \frac{1}{4} \beta^{--+} c^N\,.
\eea
The BRST operator can be split according to our gradation as
${\cal Q}_{BRST}={\cal Q}_0 + {\cal Q}_1 + {\cal Q}_2$, where
\bea
{\cal Q}_0 &=& \frac{1}{2\pi i} \oint d^2z\, [-\frac{\ka}{2} c + c^M
\mu + c^N \nu]\non\,,\\
{\cal Q}_1 &=& \frac{1}{2\pi i} \oint d^2z\, [ -\gamma^{-+-}\psi -
\gamma^{+--}(\chi-\frac{1}{4}\psib\nu) - \gamma^{+++}(\chib -
\frac{1}{4}\psib\mu)]\,,\non\\
{\cal Q}_2 &=& \frac{1}{2\pi i} \oint d^2z\,
[c(E^-+\frac{1}{2}E^-_{gh}) - c^M(M^-+\frac{1}{2}M^-_{gh}) -
c^N(N^-+\frac{1}{2}N^-_{gh}) -
\gamma^{-+-}(j^{+-+}+ \non\\
& &+\frac{1}{2}j^{+-+}_{gh}) -
\gamma^{+--}(j^{-++}+\frac{1}{2}j^{-++}_{gh}) + \gamma^{++-} j^{--+}
-
\gamma^{+++}(j^{---}+\frac{1}{2}j^{---}_{gh})]\,.
\eeal{brstN4}
Analogously to the $N\!=\!1$ string, we can now solve the BRST
cohomology and arrive at a realization of the non-linear $N\!=\!4$
superconformal algebra. Since the explicit expressions are rather
lengthy, we state only the correspondence of the $s\ell(2)$ highest
weights with the generators of the $N\!=\!4$ algebra:
\bea
\hat{E}^+ \rightarrow T &,& \non\\
\hat{M^i} \rightarrow A^{+i} &,& \hat{N^i} \rightarrow
A^{-i}\,,\non\\
\hat{j}^{+++} \rightarrow G_+ &,& \hat{j}^{+--} \rightarrow G_-
\,,\non\\
\hat{j}^{++-} \rightarrow G_{+K} &,& \psi \rightarrow G_{-K}\,.
\eea
After the Miura transformation, we get a free-field realization in
terms of the Cartan generators and the auxiliary fields. To linearize
the algebra, we adjoin four fermions and a $U(1)$ current with
operator
products
\be
\rho(z)\rhob(w) = \tau(z)\taub(w) = \frac{\ka}{(z-w)} \,,\qquad u(z)
u(w) = \frac{2 \ka}{(z-w)^2}\,.
\ee
The necessary redefinitions of the generators are given in (B.8). We
take here $(\rho,\rhob)$ corresponding to $(Q_+,Q_-)$ and
$(\tau,\taub)$ corresponding to $(Q_{+K},Q_{-K})$. The final form of
the generators can now be fixed by requiring that they match the form
of the twisted $N\!=\!4$ currents of the $N\!=\!2$ string. This leads
to the following similarity transformations:
\bea
S &=& \exp\; \Big[\frac{1}{2\pi i} \oint dz \,
(-\frac{1}{\ka}\psib\chi\mu - \frac{1}{\ka}\psib\chib\nu -
\frac{1}{2\ka}\psib \hat{M}^0\rhob - \frac{1}{\ka}\psib
\hat{N}^0\rhob + \frac{1}{4\ka}\psib u\rhob +
\frac{1}{2\ka}\psib\mu\tau + \non\\ & & + \frac{1}{2\ka}\psib\nu\taub
+ \frac{1}{\ka^2}\psib\chi\chib\rhob -
\frac{1}{4\ka^2}\psib\tau\taub\rhob + \frac{1}{4\ka}\psib\mu\mub\rhob
- \frac{1}{4\ka}\psib\nu\nub\rhob)\Big] \,,\non\\ R &=& \exp\;
\Big[\frac{1}{2 \pi i} \oint dz\,(\frac{1}{2\ka} \mub\tau\rhob -
\frac{1}{2\ka} \mub\taub\rhob)\Big]\,.
\eea
which act on the currents as follows:
\bea
\tilde{T} = R S T S^{-1} R^{-1}\,,&\non\\
\tilde{G}_a = R S G_a S^{-1} R^{-1}\,,&\tilde{A}^{\pm i} = R S A^{\pm
i} S^{-1} R^{-1}\,,\non\\
Q_+ = R S \rho S^{-1} R^{-1}\,,&Q_- = R S \rhob S^{-1}
R^{-1}\,,\non\\
Q_{+K} = R S \tau S^{-1} R^{-1}\,,&Q_{-K} = R S \taub S^{-1}
R^{-1}\,,\non\\
\tilde{u} = R S u S^{-1} R^{-1}\,.
\eeal{RSgens}
(Note that $R$ and $S$ do not commute with each other.)
Finally, we bosonize the Cartan currents
\bea
\hat{E}^0 = \frac{i\sqrt{\ka}}{4} (\del\phi_1 + \del
\phib_1)\,,\non\\
\tilde{u} =  i\sqrt{\ka} (\del\phi_2 + \del\phib_2) \,,\non\\
\hat{M}^0 = \frac{i\sqrt{\ka}}{4} (\del\phib_1 - \del \phi_1 + \del
\phi_2 - \del \phib_2)\,,\non\\
\hat{N}^0 = \frac{i\sqrt{\ka}}{4} (\del\phib_1 - \del \phi_1 - \del
\phi_2 + \del \phib_2)\,,\non\\
{\rm where}\qquad\ \ \del\phi_i(z) \del\phib_j(w) =
-\frac{\delta_{ij}}{(z-w)^2}\,,&
\eea
and rescale the fields
\bea
\rho \rightarrow \frac{1}{\sqrt{\ka}}\rho\,,&\rhob \rightarrow
\frac{1}	{\sqrt{\ka}}\rhob\,,\non\\
\chi \rightarrow \frac{2i}{\sqrt{\ka}}\chi\,,&\chib \rightarrow
\frac{2i}{\sqrt{\ka}} \chib\,,\non\\
\psi \rightarrow  \frac{1}{2} \psi \,,& \psib \rightarrow 2
\psib\,,\non\\
\mu \rightarrow  \frac{1}{2} \mu \,,& \mub \rightarrow 2
\mub\,,\non\\
\nu \rightarrow  \frac{1}{2} \nu \,,& \nub \rightarrow 2
\nub\,,\non\\
\tau \rightarrow \frac{1}{2\ka} \tau\,,& \taub \rightarrow 2 \taub\,.
\eea
The currents \equ{RSgens} then become precisely the currents of the
BRST algebra of the $N\!=\!2$ string. Since these expressions are in
general
rather long, we give here just the most interesting one:
\bea
G_{+K} &=&  \psib(-\del\phi_1\del\phib_1 -
i\frac{\sqrt{\ka}}{2}(\del^2\phi_1+\del^2\phib_1)
-\del\phi_2\del\phib_2 - \frac{1}{2}\chi\del\chib - \frac{1}{2}
\chib\del\chi -\frac{1}{2} \rho\del\rhob - \frac{1}{2} \rhob\del\rho
)-\non\\
& &- i\frac{\sqrt{\ka}}{2}\del\psib(\del\phi_2 - \del\phib_2) +
\nub(\frac{1}{2} \del\phi_1\chi + \frac{1}{2}\del\phi_2\rho + i
\frac{\sqrt{\ka}}{2}\del\chi) - i\frac{\sqrt{\ka}}{2}\del\nub\rho
+\mub(\frac{1}{2} \del\phib_1\chib -  \non\\
& & - \frac{1}{2}\del\phib_2\rhob + i \frac{\sqrt{\ka}}{2}\del\chib)
- i\frac{\sqrt{\ka}}{2}\del\mub\rhob +
\tau ( - \chib\chi - \rhob\rho +
i\frac{\sqrt{\ka}}{2}(\del\phi_1+\del\phib_1+\del\phi_2-\del\phib_2))
+\non\\
& & +  \tau\mub\mu + \frac{1}{2}\mub\del\nub\taub -
\frac{1}{2}\del\mub\nub\taub - \tau\nub\nu - \psi\mub\nub -
\psib\psi\del\psib - \psib\tau\del\taub - \frac{1}{2}\psib\mub\del\mu
- \non\\
& & -\frac{1}{2}\psib\nub\del\nu + \frac{1}{2}\psib\del\mub\mu +
\frac{1}{2}\psib\del\nub\nu -\del\psib\tau\taub -\del\psib\mub\mu +
\del\psib\nub\nu\,.
\eeal{brstN2DS}
Clearly \equ{brstN2DS} represents the BRST-operator of the $N\!=2\!$
string. The matter system is represented by $\phi_1$, $\phib_1$
and by $\chi$, $\chib$, the Liouville system by $\phi_1$, $\phib_1$
and by $\rho$, $\rhob$. The ghost system turns out to be given by
$\psi$, $\psib$ for the diffeomorphism ghosts, $\mu$, $\mub$ and
$\nu$, $\nub$ for their bosonic superpartners and $\tau$, $\taub$ for
the $U(1)$ ghosts. The background charge of the Liouville system that
cancels the conformal anomaly is given by $Q = i\sqrt{\ka}$, where,
of course, $\ka$ is the level of the $osp(4|2)$ WZW-model.
%
%
\section{$s\ell(2|1)$ Embeddings and
 the Gauge Invariant Action for General N}

We have shown in the previous chapters how one can obtain $N\!=\!1$
and
$N\!=\!2$ strings by Hamiltonian reduction. We would now like to
briefly
discuss the general properties of this procedure for $osp(N|2)$ with
arbitrary $N$.

As mentioned above, one expects that $N\!-\!2$-extended
string theories will be obtained by embeddings of $s\ell(2|1)$ into
$osp(N|2)$. Under the adjoint action of this embedding, the latter
algebra will decompose into irreducible representations of
$s\ell(2|1)$. The bosonic part of $s\ell(2|1)$ consists of an
$s\ell(2)$ algebra $e_{\pm},e_0$ and a U(1)-part $u$. The four
fermionic generators $t_{\pm}$, $\bar{t}_{\pm}$ transform as
$(\frac{1}{2},1)$ and $(\frac{1}{2},\bar{1})$, respectively:
\bea
\left.\right.[e_0,t_\pm] = \pm \frac{1}{2} t_\pm & [e_0,\bar{t}_\pm]
= \pm \frac{1}{2} \bar{t}_\pm\,,\non\\
\left.\right.[u,t_\pm] = t_\pm & [u,\bar{t}_\pm] = -  \bar{t}_\pm\,,
\eea
Irreducible representations are labelled by two quantum numbers
$(j,q)$, where $j$ is the spin and $q$ the $u$-charge. Regular
irreducible representations span a multiplet

\begin{center}
\unitlength1cm
\begin{picture}(6,4)
\put(1.6,3.5){$\left|j,q\right>$}
\put(-0.6,2.0){$\left|j-\frac{1}{2},q-1\right>$}
\put(2.5,2.0){$\left|j-\frac{1}{2},q+1\right>$}
\put(1.2,0.5){$\left|j-1,q\right>$}
\put(1.8,0.9){\line(-5,6){0.7}}
\put(2.2,0.9){\line(5,6){0.7}}
\put(1.1,2.5){\line(5,6){0.7}}
\put(2.85,2.5){\line(-5,6){0.7}}
\end{picture}
\end{center}
with $t_+ \left|j,q\right> = \bar{t}_+ \left|j,q\right> = 0$.
The embedding $s\ell(2|1)\hookrightarrow osp(N|2)$ is defined by the
bosonic $s\ell(2)$ sub-algebra and by a Cartan element of the
$so(N)$-part. For the string-related Hamiltonian reduction, we
require $q=0$ for all multiplets. This results into the following
decomposition of $osp(N|2)$ with respect to $s\ell(2|1)$:

\begin{center}
\unitlength1cm
\begin{picture}(16,5.5)
\put(1.6,4.5){$\left|1,0\right>$}
\put(0.5,3.0){$\left|\frac{1}{2},-1\right>$}
\put(2.5,3.0){$\left|\frac{1}{2},1\right>$}
\put(1.6,1.5){$\left|1,0\right>$}
\put(1.8,0.5){$(1)$}
\put(1.8,1.9){\line(-5,6){0.7}}
\put(2.2,1.9){\line(5,6){0.7}}
\put(1.1,3.5){\line(5,6){0.7}}
\put(2.85,3.5){\line(-5,6){0.7}}
\put(7.2,3.6){$\left|\frac{1}{2},0\right>$}
\put(6.1,2.1){$\left|0,-1\right>$}
\put(8.2,2.1){$\left|0,1\right>$}
\put(7,0.5){$(N-2)$}
\put(6.7,2.5){\line(5,6){0.7}}
\put(8.5,2.5){\line(-5,6){0.7}}
\put(13,3.0){$\left|0,0\right>$}
\put(11.5,0.5){$(\frac{1}{2}N(N-1)-2N+3)$}
\end{picture}
\end{center}

The numbers below the multiplets denote the multiplicities. Of
course, the $(j\!=\!1)$-multiplet corresponds precisely to the
$s\ell(2|1)$ itself. It is fairly obvious that the string BRST
current and the anti-ghost will arise from this multiplet, if we
choose the gradation corresponding to $(e_0+u)$. From this it becomes
clear that the U(1) part $u$ will represent the ghost number current
of the string model. If we had chosen an embedding leading to
multiplets with $q\ne0$, these multiplets would have non-vanishing
ghost number, and this appears not acceptable for a string theory.

The $N-2$ multiplets with $j=\frac{1}{2}$ will give rise to the
underlying $N-2$ supersymmetries of the string. The states
$\left|0,-1\right>$ and $\left|0,1\right>$ belong to the $so(N)$ part
of $osp(N|2)$. Note that there are in general $s\ell(2)$ highest
weights that are negatively graded. These will give rise to bosonic
auxiliary fields which correspond to the $N-2$ $\beta,\gamma$ systems
of the string. Note also that the $j=\frac{1}{2}$ multiplets are
short multiplets, which typically arise in the non-linear form of
superconformal algebras.

Simple counting shows that we will have in addition
$\frac{1}{2}N(N-1)-2N+3$ singlets. They are simply the rest of the
generators of the $so(N)$ part. We note that due to this there will
be an additional difficulty in the reduction for $N\ge5$. Namely
there will be generators of the $so(N)$ part besides the Cartan
elements that do not get automatically bosonized. Nevertheless, from
cursory inspection we expect that one can find a bosonization of them
that allows for a string interpretation.

In order to find gauge invariant lagrangian for general $osp(N|2)$,
let us first make a remark concerning different definitions of
gradations. In the reduction process it is convenient to think of the
gradation being defined on the current algebra. Thus the ghosts
acquire non-trivial grades via to the ghost contribution to the
currents, whereas the auxiliary fields have grade zero. This is
important for finding the filtrations of BRST operators
\equ{filtration},\equ{brstN4}. However, for finding the gauge
invariant
lagrangian, it is more convenient to define the gradation on the
algebra itself. That is, we will assign non-trivial grades also
to the auxiliary fields.

The starting point is the action
\be
S_0 =-\frac{1}{2\pi} \int d^2z\, str(\frac{\ka}{2}\Ab e_- + \Ab \psi
+ \Ab \varphi )\,.
\eel{S0}
This is the a priori non-invariant part of the WZW-model on
$osp(N|2)$, subject to the constraint that the lowering operator of
$s\ell(2)$ is equal to $\frac{\ka}{2}$. The field $\psi$ is the
auxiliary field for the negatively graded highest weight of
$s\ell(2|1)$ and $\varphi$ are the auxiliary fields for the
negatively graded $so(N)$ currents.

Under a gauge transformation we have ($\eta$ being the gauge
parameter)
\bea
\delta S_0 &=& -\frac{1}{2\pi} \int d^2z\,  str \Big(
\Ab_\frac{1}{2}(\frac{\ka}{2}[\eta_\frac{1}{2},e_-] +
[\eta_\frac{1}{2},\varphi_{-1}]) + \delb \eta_\frac{1}{2}
\psib_{-\frac{1}{2}} + \delb \eta_1 \varphi_{-1} + \non\\ & &
\qquad\qquad\qquad\qquad  +\Ab_\frac{1}{2}\delta\psi_{-\frac{1}{2}} +
\Ab_1 \delta\varphi_{-1}\Big)\,.
\eea
The subscripts denote explicitly the grades of the fields. Since
under the decomposition $osp(N|2)\rightarrow s\ell(2|1)$ the
half-integer graded fields can come only from the fermionic part, we
can decompose $\Ab_\frac{1}{2}$ uniquely into $\Ah_\frac{1}{2} \in
ker~ad_{e_+}$ and $\At_\frac{1}{2} \in ker~ad_{e_-}$. Let us now look
at the term of the form $str(\Ah_\frac{1}{2}
[\eta_\frac{1}{2},\varphi_{-1}])$. We see that $[\eta_\frac{1}{2},
\varphi_{-1}]$ has to be an element of $ker~ad_{e_-}$ to have
non-vanishing supertrace. Since $\varphi_{-1}$ lies in both
$ker~ad_{e_+}$ and $ker~ad_{e_-}$, this shows that only the part of
$\eta_\frac{1}{2}$ contributes to this term which lies in
$ker~ad_{e_+}$. We denote this by $\etat_\frac{1}{2}$. Similar
arguments lead us to introduce $\etah_\frac{1}{2} \in ker~ad_{e_-}$.
We arrive thus at
\bea
\delta S_0 &=& -\frac{1}{2\pi} \int d^2z\, str \Big(
\Ah_\frac{1}{2}(\frac{\ka}{2}[\etat_\frac{1}{2},e_-] +
[\etah_\frac{1}{2},\varphi_{-1}]) +
\At_\frac{1}{2} ([\etat_\frac{1}{2},\varphi_{-1}] +
 \delta\psi_{-\frac{1}{2}}) +\non\\
& & \qquad\qquad\qquad\qquad + \delb \etah_\frac{1}{2}
\psi_{-\frac{1}{2}} + \delb \eta_1\varphi_{-1} +
\Ab_1\delta\varphi_{-1} \Big)\,.
\eeal{deltaS0}
It is easy to see that gauge invariance can be restored by modifying
the constraints, that is, by adding the following term to the
lagrangian:
\be
S_1  = -\frac{1}{2\pi} \int d^2z\, str(
\At_\frac{1}{2}([\psib_\frac{1}{2},\varphi_{-1}] + [\chi,e_-]) )\,,
\ee
where the new field $\chi$ lies in the part of $osp(N|2)$ with
half-integer grade. $\psib_\frac{1}{2}$ is conjugate to
$\psi_{-\frac{1}{2}}$ and therefore lies in $ker~ad_{e-}$. Gauge
invariance also forces us to introduce kinetic terms for the
auxiliary fields,
\be
S_{kin} = -\frac{1}{2\pi} \int d^2z\, str(
\delb\psib_\frac{1}{2}\psi_{-\frac{1}{2}} +
\delb\bar{\varphi}_1\varphi_{-1} + \frac{1}{\ka}\delb\chi[\chi,e_-]
)\,.
\eel{Skin}
The whole action is gauge invariant if we assign to the
auxiliary fields the following transformation rules:
\bea
\delta\varphi_{-1} = 0 &,&\delta\bar{\varphi}_1 = \eta_1 -
[\etat_\frac{1}{2},\psib_\frac{1}{2}]\,,\non\\
\delta\psi_{-\frac{1}{2}} = \etah_\frac{1}{2} &,&
\delta\psi_{-\frac{1}{2}} =
-[\etat_\frac{1}{2},\varphi_{-1}]\,,\non\\
\delta\chi = \frac{\ka}{2} \etat_\frac{1}{2}\,.
\eea
{}From the gauge invariant action we can proceed in complete analogy
to the previous sections, and infer the form of the constraints. The
interpretation of the auxiliary fields is clear, in that
$(\psi,\psib)$ will always correspond to the diffeomorphism ghosts of
the string. Their bosonic superpartners can be identified with
$(\varphi,\bar{\varphi})$. The additional ghosts that are present in
string theories with extended supersymmetries will come in by
linearizing the algebra. From our experience with the $N\!=\!1$ and
$N\!=\!2$
strings we expect $\chi$ to describe the fermions in the matter
sector.
%
%
\section{A Note on Topological Strings}
%
%
As pointed out, twisted $N\!=\!2$ algebras formed by the
BRST-current, the anti-ghost together with the energy momentum tensor
and the ghost current are one of the most basic features of the BRST
formulation of string theories. The construction of topological
strings \cite{v2} on the other hand involves from the very beginning
realizations of twisted $N\!=\!2$ superconformal algebras for the
matter, Liouville and ghost sector separately. The BRST operator of
topological strings has the special property that it is the sum of
${\cal Q}_s = \frac{1}{2\pi} \oint G_+ dz$ and ${\cal Q}_v =
\frac{1}{2\pi} \oint J^v_{BRST} dz$, where $G_+$ denotes the sum of
the supercharges of the matter, Liouville and ghost systems and
$J^v_{BRST}$ is the BRST current that arises by gauge fixing the
local world sheet symmetries. According to our general reasoning, we
expect also for the topological string that $J^v_{BRST}$ and $b$ form
a twisted $N\!=\!2$ superconformal algebra. This suggests that there
should actually be a doubly twisted $N\!=\!4$ superconformal algebra
in the topological string.

Let us first state some remarks on this hidden, doubly twisted
$N\!=\!4$ superconformal algebra. We focus on the twisted currents
\bea
\Th &=& T + \del M_{+-} + \del M_{+K-K} + \frac{1}{2} \del U = T + i
\del A^{+3}
 +  \frac{1}{2} \del U ,\non\\
\Gh_{+} &=& G_{+} + \del Q_{+} ,\non\\
\Gh_{-} &=& G_{-} + \del Q_{-} ,\non\\
\Gh_{+K} &=& G_{+K} - \del Q_{+K} ,\non\\
\Gh_{-K} &=& G_{-K} + \del Q_{-K} ,\non\\
\Jh_1 &=& 2 i A^{+3} ,\non\\
\Jh_2 &=& i ( A^{+3} - A^{-3}) + U\ .
\eeal{doubletwist}
Note that the twist assigns conformal weight $1$ to $\{G_+,G_{+K}\}$,
weight $2$ to $\{G_-,G_{-K}\}$, wight $0$ to $A^{++}$, weight $2$ to
$A^{+-}$. The U(1)-currents and $A^{--}$ of the sub-algebra stay at
conformal weight $1$.

At $k^+ = k^-$ the currents $ \{\Th,\Gh_{+},\Gh_{-},\Jh_1\} $ form a
topologically twisted $N\!=\!2$ sub-algebra with central extension
$6k^+$,
whereas $ \{\Th,\Gh_{+K},\Gh_{-K},\Jh_2\} $ form a topologically
twisted $N\!=\!2$ sub-algebra with vanishing central extension. It is
quite remarkable that \equ{doubletwist} together with
$\{A^{++},A^{+-},A^{--},Q_-,Q_{+K}\}$ form a closed sub-algebra. In
the following we will see that it is exactly this sub-algebra that is
realized in the topological string.

In the topological string we take the matter
sector to be completely arbitrary. We only require that it exhibits a
twisted $N\!=\!2$ superconformal symmetry, with some (twisted)
central extension $c_m$. In representing the Liouville sector we
follow \cite{v2} and introduce two bosonic fields $\{\pi,\varphi\}$
and two fermions $\{\chi,\psi\}$ with OPE's
\bea
\pi(z)\varphi(w) = -\ln(z-w) \,,&& \del\chi(z)\psi(w) = \frac{1}{z-w}
{}.
\eea
The generators of the twisted $N\!=\!2$ superconformal algebra in the
Liouville sector are then
\bea
J_l &=& - \del\chi\psi - Q \del\varphi + Q \del\pi ,\non\\
G_{l+} &=& \del\pi \psi - Q \del \psi ,\non\\
G_{l-} &=& -\del\chi\del\varphi + Q \del^2 \chi ,\non\\
T_l &=& -\del\pi\del\varphi -\del\chi\del\psi + Q\del^2\pi\ ,
\eea{}
where $Q$ is an arbitrary background charge, in terms of which the
central extension is $c_l = 3 + 6 Q^2$.
Since only the $G_-$ supercharge corresponds to a local gauge
symmetry, we have for the ghosts the usual fermionic $(b,c)$ system
for the diffeomorphisms, plus a bosonic $(\beta,\gamma)$ system as
their superpartners, with
\be
b(z)c(w) = \beta(z)\gamma(w) = \frac{1}{z-w}
\ee
(both ghost systems have spins $(2,-1)$).
The $N\!=\!2$ generators in the ghost sector look
\bea
J_g &=&  2 \beta\gamma - bc ,\non\\
G_{g+} &=& -b\gamma ,\non\\
G_{g-} &=& -\del\beta c - 2 \beta\del c ,\non\\
T_g &=& -2 b\del c - \del b c + 2 \beta \del\gamma +
\del\beta \gamma \ ,
\eea{}
and give rise to a twisted central extension of $c_{g}=-9$.

We are free to add the following total derivatives to
the BRST current,
\bea
J^v_{BRST} &=& c(T_m + T_l + \frac{1}{2}T_g) + \gamma (G_{m-} +
G_{l-} +
 \frac{1}{2} G_{g-}) - \non \\&&(Q+\frac{c_m}{6 Q})\del(c\del\pi +
\gamma\del\chi) \,,
\eeal{brsttop}
and introduce the ghost current:
\be
I_{g} = -bc + \beta\gamma + (Q+\frac{c_m}{6 Q})\del\pi \ .
\ee
The improvement terms in $J^v_{BRST}$ and in $I_{g}$ are such that
$\{T_{tot},J^v_{BRST},b,I_{gh}\}$ indeed forms a twisted $N\!=\!2$
superconformal algebra\foot{This algebra with vanishing central
charge is of course different and independent from the
``constituent'' $N\!=\!2$ algebras and is analogous to the $N\!=\!2$
algebra that arises in the bosonic string.} with vanishing
central extension. Furthermore, we find that
\bea
A^{++} &=& -i c(G_{m+} + G_{l+} +\frac{1}{2} G_{g+}) + i \gamma(J_m +
J_l +\frac{1}{2} J_g)-i\, k\,\del\gamma ,\non\\
A^{+3} &=& -\frac{i}{2} (J_m + J_l + J_g) ,\non\\
A^{+-} &=& -i \beta\ .
\eeal{app}
form an $s\ell(2)$ Kac-Moody algebra at level
\be
k\ =\ \frac{1}{6}\,c_{tot}\ ,
\eel{klevel}
where
\be
c_{tot} \equiv c_{l}+c_{g}+c_{m} =  6(Q^2 -1) + c_m\ .
\eel{ctot}
Together with
\bea
A^{--} &=& i c (G_{m-} + G_{l-} + \frac{1}{2} G_{g-})
+ i (Q+\frac{c_m}{6Q})\del(\del\chi c) ,\non\\
Q_- &=& \frac{1}{2} \beta c + \frac{1}{2} (Q+\frac{c_m}{6Q}) \del\chi
,\non\\
Q_{+K} &=& - \frac{1}{2} c( J_m + J_l + \frac{1}{2} J_g) +
\frac{1}{2}
(Q+\frac{c_m}{6Q})(\del\chi\gamma + \del\pi c)
+\frac{1}{2}k\,\del c\
\eea{}
we thus see that the above-mentioned sub-algebra of the hidden,
doubly
twisted $N\!=\!4$ algebra is indeed realized in the topological
string, with
\be
k^+\ = k^-\ =\ k\ .
\ee
Note that due to the topological nature of the theory, the BRST
operator $\oint J^v_{BRST}$ is nilpotent for any $c_{tot}$, and thus
there is, strictly speaking, no critical central charge for the
topological string. Nevertheless, the ``critical'' case $c_{tot}=0$
is distinguished in the BRST algebra in that the level \equ{klevel}
of the $s\ell(2)$ algebra vanishes and the $s\ell(2)$ raising
generator $A^{++}$ \equ{app} decouples from the gravitational
descendant,
$\del\gamma$.

Note also that the similarity transformation $U$ that rotates the
topological string into the matter picture \cite{ekyy} arises here
naturally from a Kac-Moody current:
\be
U\ = \exp\,\big[\frac{1}{2\pi}\oint dz\,A^{--}\big]\ .
\ee
It is thus nothing but an inner automorphism of the doubly twisted
$N\!=\!4$ superconformal algebra.

Although the structure of the doubly twisted $N\!=\!4$ algebra seems
to be superficially similar to what we found for the $N\!=\!2$
string, there is an important difference. We do not get here enough
fermions such as to form the non-linear version of the algebra, and
it is precisely the non-linear form that arises from Hamiltonian
reduction of $osp(4|2)$. Therefore it is unclear at the moment how
one can construct topological strings along the lines that we
presented in the previous sections. We will leave this problem for
future investigation.
%
%
\section{Conclusions}
%
%
The presented results clearly point to a systematic treatment of
crticial and non-crititcal string theories. Indeed the structure
revealed in the studied cases suggests that a very large class of
string theories could be obtained from those Hamiltonian reductions
which give rise to extensions of the $N=2$ superconformal algebras.
Such reductions correspond to embeddings of $s\ell(2|1)$ in super Lie
algebra. In addition, the grading has to be chosen in such a way that
the realization obtained is consistent with the BRST structure,
\equ{QG}.

Though the complete classification of $sl(2|1)$ embeddings in Lie
superalgebras has been achieved \cite{RaSo}, this program is not yet
finished. All examples studied so far have as a common characteristic
that in the decomposition of the adjoint representation of the super
Lie algebra in terms of irreducible representations of the embedded
$sl(2|1)$, only typical representations with vanishing $U(1)$ charge
occur. Exactly for such cases we believe that the presented
techniques are sufficient to recover the non-critical string theory.
On the other hand, for generic $sl(2|1)$ embeddings, non-trivial
$U(1)$ charges and atypical representations can occur. At this moment
it is not clear how to generalize the methods developed in this paper
to cover such cases. Further work in this direction is under way.

Several other questions remain unanswered. We found a hidden, doubly
twisted $N=4$ structure in topological gravity. An obvious question
is whether topological gravity can be obtained from Hamiltonian
reduction. This is a relevant question, as the construction of
extensions of topological gravity, such as topological $W$ gravity,
turns out to be technically extremely involved. A more systematic
approach, based on Hamiltonian reduction would greatly facilitate
this task.
%
%
\newpage
\begin{appendix}
\section{Conventions for $osp(N|2)$}
We summarize our conventions for the current algebra of $osp(N|2)$:
\bea
E^0(z) E^0(w) &=& \frac{\ka}{8} \frac{1}{(z-w)^2}\,,\non\\
E^+(z) E^-(w) &=& \frac{\ka}{4} \frac{1}{(z-w)^2} +
\frac{E^0(w)}{(z-w)}\,,\non\\
E^0(z) E^\pm(w) &=& \pm\frac{1}{2}\frac{E^\pm(w)}{(z-w)}\,,\non\\
\non\\
J^m(z) J^n(w) &=&= \frac{1}{8}\frac{\delta^{mn}}{(z-w)^2} -
\frac{1}{4}\frac{\sqrt{2}f_{mn}\:^k J^k(w)}{(z-w)}\,,\non\\
\non\\
E^0(z) j^{\pm k}(w) &=& \pm \frac{1}{4}\frac{j^{\pm
k}(w)}{(z-w)}\,,\non\\
E^\pm(z) j^{\mp k}(w)&=& \frac{1}{2} \frac{j^{\pm
k}(w)}{(z-w)}\,,\non\\
\non\\
J^m(z) j^{\pm k}(w) &=&\frac{1}{4} \frac{\sqrt{2}\la_{mn}\:^k\,
j^{\pm n}(w)}{(z-w)}\,,\non\\
\non\\
j^{+m}(z) j^{-n}(w) &=& \frac{\ka}{8} \frac{\delta^{mn}}{(z-w)^2} +
\frac{1}{4} \frac{\delta^{mn}\, E^0(w)}{(z-w)} +
\frac{1}{4}\frac{\sqrt{2} \la_{mn}\:^k J^k(w)}{(z-w)}\,,\non\\
j^{\pm m}(z) j^{\pm n}(w) &=& \mp\frac{1}{4} \frac{\delta^{mn}\,.
E^{\pm}}{(z-w)}\,.
\eea
Here $E^0$, $E^\pm$ denote the $s\ell(2)$ part, $J^m$ the $so(N)$
part and $j^{\pm k}$ are the fermionic currents. For the $so(N)$
algebra we employ a double index notation where $m$ stands for $pq$
with $1\le p < q \le N$. The structure constants are given by
$f_{mn}\:^k = [\la^m , \la^n]$ with $\la^{pq}\:_{rs} =
\frac{1}{\sqrt{2}} (\delta^p_r\delta^q_s - \delta^p_s\delta^q_r)$.
The metric used in our computations is
\be
g_{+-} = -2\,,\qquad g_{00} = -4\,,\qquad g_{+m-n} =
-4\delta_{mn}\,,\qquad g_{mn}= -4\delta_{mn}\,.
\ee
%
%
\section{Superconformal algebras with $N\!=\!3,4$}
\begin{enumerate}
\item{$N\!=\!3$}\\
The $N\!=\!3$ superconformal algebra consists of the stress tensor,
three supercharges, an $so(3)$ Kac Moody algebra at level $k$ and an
additional fermionic current $F$ with spin $\frac{1}{2}$. All the
fields are primary with respect to the stress tensor. The
non-vanishing operator products are
\bea
G^a(z) G^b(w) &=& \frac{2 g^{ab} k}{(z-w)^3} + \frac{\eps^{ab}\;_c\,
K^c(w)}{(z-w)^2} + \frac{2g^{ab}\, T(w) + 2 \eps^{ab}\;_c\, \del
K^c(w)}{(z-w)}\,,\non\\
K^a(z) K^b(w) &=& \frac{-g^{ab} k}{(z-w)^2} + \frac{\eps^{ab}\;_c\,
K^c(w)}{(z-w)}\,,\non\\
K^a(z) G^b(w) &=& \frac{- F(w)}{(z-w)^2} + \frac{\eps^{ab}\;_c\,
G^c}{(z-w)(w)}\,,\non\\
G^a(z) F(w) &=& \frac{-K^a(w)}{(z-w)}\,,\non\\
F(z) F(w) &=& \frac{- k}{(z-w)}\,.
\eea
The components of the metric are $g^{+-} = 1$ and $g^{33} = 2$. The
central charge is given by $c = 3k$. The fermion can be factorized by
redefining the generators as follows:
\bea
\tilde{T} &=& T - \frac{1}{2k} F\del F \,,\non\\
\tilde{G^a} &=& G^a - \frac{1}{k} F K^a\,.
\eea
This then leads to the non-linear form of the algebra.

\item{$N\!=\!4$}\\
The large $N\!=\!4$ superconformal Algebra consists of the stress
tensor, four supercharges, two commuting $sl(2)$ Kac Moody algebras
at different levels $k^+,k^-$, four fermionic currents with spin
$\frac{1}{2}$ and a U(1) current. All the fields are primary, the
rest of the non-vanishing operator products are
\bea
G_a(z)G_b(w) &=& \frac{2c}{3}\frac{\delta_{ab}}{(z-w)^{3}} +
\frac{2\, M_{ab}(w)}{(z-w)^{2}} + \frac{2\,T(w)\, \delta_{ab} + \del
M_{ab}(w)}{(z-w)} \,,\non\\ \\ A^{\pm i}(z)G_a(w) &=&
\frac{\alpha^{\pm i}\,_a\,^b\, G_b(w)}{z-w} \mp
\frac{\frac{2k^{\pm}}{k}\alpha^{\pm i}\,_a\,^b\,
Q_b(w)}{(z-w)^2}\,,\non\\ A^{\pm i}(z)A^{\pm j}(w) &=&
\frac{f^{ij}\:_k\,A^{\pm k} (w)}{(z-w)} -
\frac{\frac{k^{\pm}}{2}\delta^{ij}}{(z-w)^2}\,,\non\\ Q_a(z)G_b(w)
&=& \frac{2(\alpha^{+i}_{ab}\,A^+_i(w) - \alpha^{-i}_{ab}\,A^-_i(w))
+ \delta_{ab}U(w)}{(z-w)}\,,\non\\ A^{\pm i}(z)Q_a(w) &=&
\frac{\alpha^{\pm i}\,_a\,^b\,Q_b(w)}{(z-w)}\,,\non\\ U(z)G_a(w) &=&
\frac{Q_a(w)}{(z-w)^2}\,,\non\\ Q_a(z)Q_b(w) &=&
-\frac{k}{2}\frac{\delta_{ab}}{(z-w)}\,,\non\\ U(z)U(w) &=&
-\frac{k}{2}\frac{}{(z-w)^2}\,,
\eea{}
with central charge $ c=\frac{6k^+k^-}{k} , \; \; k=k^+ +k^-  , $
and
\be
M_{ab}(z) = -\frac{4}{k}(k^- \alpha^{+i}_{ab}\,A^{+}_i(z) + k^+
\alpha^{-i}_{ab}\,A^{-}_i(z)) .
\ee
The index $i$ takes the values $\{+,-,3\}$ and $\delta_{ij}$ and
$f^{ij}\:_k $ are defined as
\bea
\delta_{+-} &=& \frac{1}{2} ,\non \\
f^{+-}\:_3 &=& -2i ,\non \\
f^{3\pm}\:_{\pm} &=& \mp i .
\eea{}
The non zero values for $\delta_{ab}$ and $\alpha^{\pm
i}_{ab}=-\alpha^{\pm i}_{ba}$ with $a,b$ running over \linebreak
$\{+,-,+K,-K\}$ are given by
\bea
\delta_{+-} = \frac{1}{2} , && \delta_{+K-K} = \frac{1}{2} ,\non \\
\alpha^{\pm 3}_{+-} = - \frac{i}{4} , && \alpha^{\pm 3}_{+K-K} = \mp
\frac{i}{4} ,\non \\
\alpha^{+-}_{++K} = \frac{i}{2} , && \alpha^{++}_{--K} = -\frac{i}{2}
,\non \\
\alpha^{-+}_{-+K} = -\frac{i}{2} , && \alpha^{--}_{+-K} = \frac{i}{2}
{}.
\eea{}
To get the non-linear form, one factorizes the fermions and the
U(1) current by redefining the generators as follows:
\bea
\tilde{T} &=& T + \frac{1}{k} (UU + \del Q^a Q_a)\,,\non\\
\tilde{G_a} &=& G_a + \frac{2}{k} U G_a - \frac{2}{3k^2}
\eps_{abcd}Q^bQ^cQ^d +
\frac{4}{k}Q^b(\alpha^{+i}_{ba}\tilde{A}^+_i -
\alpha^{-i}_{ba}\tilde{A}^-_i)\,\non\\
\tilde{A}^{\pm i} &=& A^{\pm i} - \frac{1}{k} \alpha^{\pm
i}_{ab}Q^aQ^b\,.
\eea
The four index antisymmetric tensor is given by
\be
\eps^{abcd} = \pm(4\alpha^{\pm i ab}\alpha^\pm\,_i\,^{cd} -
\delta^{ac}\delta^{bd} + \delta^{ad}\delta^{bc})\,.
\ee
\end{enumerate}
\end{appendix}
%
%

\end{document}